\documentclass[sigconf,screen]{acmart}

\usepackage{graphicx}
\usepackage{setspace}               
\usepackage{multicol}               
\usepackage{xcolor}
\usepackage{caption}
\usepackage{subcaption}
\usepackage{csquotes}

\usepackage[utf8]{inputenc}
\usepackage{url}
\usepackage{hyperref}
\usepackage{amsfonts}
\usepackage{float} 
\usepackage{listings}
\usepackage{color}
\usepackage{fancyhdr} 
\usepackage{bibentry} 
\usepackage{enumitem}
\usepackage{multirow}
\usepackage{geometry}
\usepackage{dblfloatfix}

\usepackage{array}

\usepackage{dirtytalk}

\usepackage{soul}

\definecolor{edit}{HTML}{000000}
\definecolor{minor}{HTML}{000000}

\fancypagestyle{firstpage}
{
    \fancyhead[L]{Workshop paper presented at \href{https://tominhai-cui2025.github.io/}{ToMinHAI at CUI’2025: Theory of Mind in Human-CUI Interaction}, held in conjunction with the \href{https://cui.acm.org/2025/}{2025 ACM conference on Conversational User Interfaces}, July 8th, 2025.}    
    \fancyhead[R]{}
}

\AtBeginDocument{%
  \providecommand\BibTeX{{%
    \normalfont B\kern-0.5em{\scshape i\kern-0.25em b}\kern-0.8em\TeX}}}

\setcopyright{none}           
\acmConference[ToMinHAI at CUI’2025]{ToMinHAI at CUI’2025: Theory of Mind in Human-CUI Interaction}{July 8, 2025}{Waterloo, ON, Canada}
\acmDOI{}      
\acmISBN{}     
\acmPrice{}    

\begin{document}

\title[Theory of Mind and Self-Disclosure to CUIs]{Theory of Mind and Self-Disclosure to CUIs}

\author{Samuel Rhys Cox}
\email{srcox@cs.aau.dk}
\orcid{0000-0002-4558-6610}
\affiliation{%
  \institution{Aalborg University}
  \city{Aalborg}
  \country{Denmark}
}




\begin{abstract}
Self-disclosure is important to help us feel better, yet is often difficult.
This difficulty can arise from how we \textit{think} people are going to react to our self-disclosure.
In this workshop paper, we briefly discuss self-disclosure to conversational user interfaces (CUIs) in relation to various social cues. 
We then, discuss how expressions of uncertainty or representation of a CUI's reasoning could help encourage self-disclosure, by making a CUI's intended ``theory of mind'' more transparent to users.
\end{abstract}



\keywords{Chatbots, Self-Disclosure, Theory of Mind, Reasoning and Thought, Uncertainty}

\maketitle

\thispagestyle{firstpage}

\section{Introduction}

Self‐disclosure (the revealing of personal information such as thou\-ghts, feelings, and opinions~\cite{archer1980effects}) has been shown to benefit both emotional and physical well‐being: it can reduce stress, promote self‐reflection, and encourage social support~\cite{greene2006self}.
However, we may not always feel confident and comfortable when self-disclosing if we perceive the \textit{risks} of disclosing as outweighing any potential \textit{benefits} of self-disclosure~\cite{altman1973socialpenetrationtheory,perlman1987revelation}.

These perceived risks may stem from how we \textit{think} our interlocutor will react. For example, we may think that they will judge or stereotype us, or that they will not be receptive to our disclosures~\cite{vogel2003seek}.
These form the \textit{Theory of Mind} (ToM) that we hold for others. That is: our ability to attribute mental states such as beliefs, desires, intentions, and emotions to others.

In this workshop paper, we will discuss how ToM can help us design conversational user interfaces (CUIs) as avenues of self-disclosure.
CUIs (such as text-based conversational agents (CAs) or voice-based assistants) are increasingly used by people as a means of companionship and emotional support~\cite{jo2024understanding,croes2021can}, and in April 2025 Harvard Business Review found therapy and companionship to be the number one use of generative AI~\cite{HBR-GenAI-Top10}.
Motivated by this, we discuss prior work investigating self-disclosure to CUIs. 
Here, we explore how the social cues of a CUI (such as verbal, visual, auditory, or invisible cues~\cite{FEINE2019138}) can be varied, and how this may affect people's ToM that results from perceived risks of self-disclosure.
We then discuss how a CUI's ToM could be transparently shared to users to lessen feelings of judgement, and encourage self-disclosure.


\section{CUIs for Self-Disclosure}

The social cues of a CUI can be adapted to encourage people's self-disclosure. 
A common thread among some prior work is that these social cues can be adapted to reduce people's \textit{feelings of judgement} from the CUI. E.g., people's beliefs that:
\begin{displayquote}
    \centering
    ``\textit{I think that this chatbot will judge me}''\\
    vs.\\
    ``\textit{I think that this chatbot will \textbf{not} judge me}''
\end{displayquote}
For example, studies have found that people may feel less judged when disclosing to a CUI compared to a human~\cite{jacobsen2025chatbots,pickard2016revealing}, with common threads of research connecting the level of social presence to feelings of judgement (e.g., people \textit{may} feel more judged with higher levels of social presence, such as more anthropomorphised visual cues~\cite{PICKARD2020106197}, or precise recall of past user utterances~\cite{cox2023comparing}).
Demonstrating this, Pickard et al. found that people were more likely to disclose shame to a disembodied voice agent compared to both an embodied CA or a human interviewer~\cite{PICKARD2020106197}.
Similarly, Wang et al. found people were more willing to disclose to an ``\textit{algorithm}'' than a person or AI assistant~\cite{yanyun2024recommendationtrust}.

Studies have also investigated how changing verbal cues in both terms of conversational content~\cite{cox2023comparing,jacobsen2025chatbots,cox2025ephemerality} and conversational style~\cite{cox2022does,chen2024different} can affect self-disclosure.
For example, studies have investigated the effect of reciprocal self-disclosure (motivated by Social Penetration Theory~\cite{altman1973socialpenetrationtheory}) whereby people disclose more if a chatbot itself discloses to them~\cite{lee2020hear,adam_onboarding_2019,ravichander2018empirical,moon2000intimate}.
Recent work has also investigated the effect of multiple social cues (such as interaction effects of both visual and verbal cues~\cite{chen2024different}) reflecting the breadth and complication of potential designs.

Finally, the beliefs of the user themselves effect people's likelihood to self-disclose.
For example, Sundar and Kim found that as people's belief in the machine heuristic (a belief that machines are more secure and trustworthy than humans) increased, so too did their likelihood to disclose personal information to a CUI~\cite{sundar2019machine}.
Additionally, our own work has found that people's views of chatbots as either a tool or a conversational partner affects preferred social cues~\cite{cox2023comparing}.
Further, people's pre-existing beliefs regarding CUI emotional capacity and intelligence impact cue effectiveness~\cite{cox2022does,liu2018should}.


While some results may seem somewhat contradictory 
(e.g., some studies have conflicting results regarding the impact of a CUI's social presence on self-disclosure)
we draw on the common thread that people often do not self-disclose due to perceived risks and feelings of judgement from CUIs, as well as apprehension to self-disclose often being mediated by people's personal beliefs.



\section{Expressions of Uncertainty}

Prior research and theories have posited that people generally dislike uncertainty in communication, and that more concrete and unambiguous information will drive communication~\cite{berger1974some}. 
Connected to this, some barriers to self-disclosure are related to the uncertainty regarding the beliefs of your interlocutor (e.g., ``\textit{is my interlocutor likely to stigmatise me?}'').
Uncertainty such as this can produce anxiety and disrupt self-disclosure.
For example, Sien and McGrener investigated people sharing stories about their mental health within a community of users. They found that people (who initially thought that others would judge or stigmatise them) felt more relieved of anxiety and open to social support after reading concrete stories from others~\cite{sien2025gentle}.

However, human-CUI interactions (like human-human interactions) may be operating on levels of incomplete information~\cite{de2017negotiating}.
This could lead either the CUI or the human user to incorporate false beliefs in their ToM.
False beliefs in each interlocutor's ToM could then lead to misunderstandings, frustration, and feelings of not being understood or appreciated.
In natural language conversations, an interlocutor could simply transparently state that they have a level of uncertainty in communication, and therefore seek clarification. Yet, depending on context, different response styles could lead to potentially negative user perceptions~\cite{wester2024ai}).

\begin{figure}[htbp]
  \centering
  \includegraphics[width=0.35\textwidth]{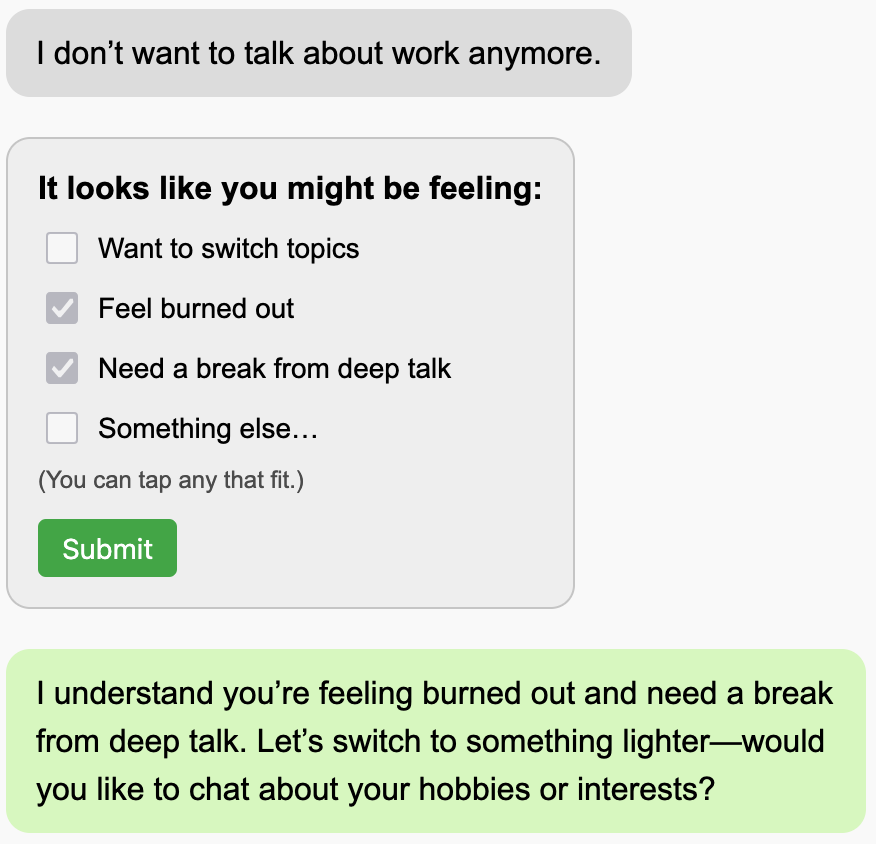}
  \caption{CUIs have different affordances to human-human conversations to express uncertainty. For example, the ability to use full natural language responses, multiple choice responses, or visual cues.}
  \label{fig:ExpressingMultiple}
\end{figure}

In addition to increasingly capable natural language responses (given the capabilities of LLMs), CUIs offer the affordance to reply using a variety of cues not available in normal human-human conversations.
For example, uncertainty could be expressed visually via changing background colours of the CUI (e.g., with colour attribution to indicate less confidence and understanding in the CUI's ToM).
Additionally, (while perhaps seeming less exciting than a fully open-ended natural language LLM-driven conversation) a CUI could seek clarification or express uncertainty using other means such as multiple choice responses (see Figure~\ref{fig:ExpressingMultiple}).






While use of multiple choice responses in Figure~\ref{fig:ExpressingMultiple} may seem like a bit of a ``\textit{step back}'' for CUIs (with this Figure offered perhaps as more of a provocation), some recent work indicates that people may not want more capable and personalised CUI responses when they are disclosing sensitive emotional information~\cite{cox2025ephemerality,park2021wrote,cox2023comparing}.
For example, in our recent work people who talked to chatbot that did not have memory between chatting sessions described not feeling judged during interactions (e.g., one user stated: ``\textit{I didn’t feel judged} [...] \textit{I felt like I was just journaling with helpful prompts being offered along the way}'')~\cite{cox2025ephemerality}.





\section{Exposing the CUI's Internal ToM}

When interacting with CUIs (such as ChatGPT) they may display ``reasoning'' to users while responses are being processed.
Such reasoning can form explanations of a CUI's intent, and could transparently share the CUI's ToM with the user. 
For example, a CUI could could directly refer to the user's fear of judgement in its reasoning before providing a response:

\begin{displayquote}
``\textit{I think the user may have an emotional need that they wish to discuss. I will not judge them, and do not want them to think I would react negatively to what they will say. When replying to the user, I should...}''
\end{displayquote}

\begin{figure}[htbp]
  \centering
  \includegraphics[width=0.5\textwidth]{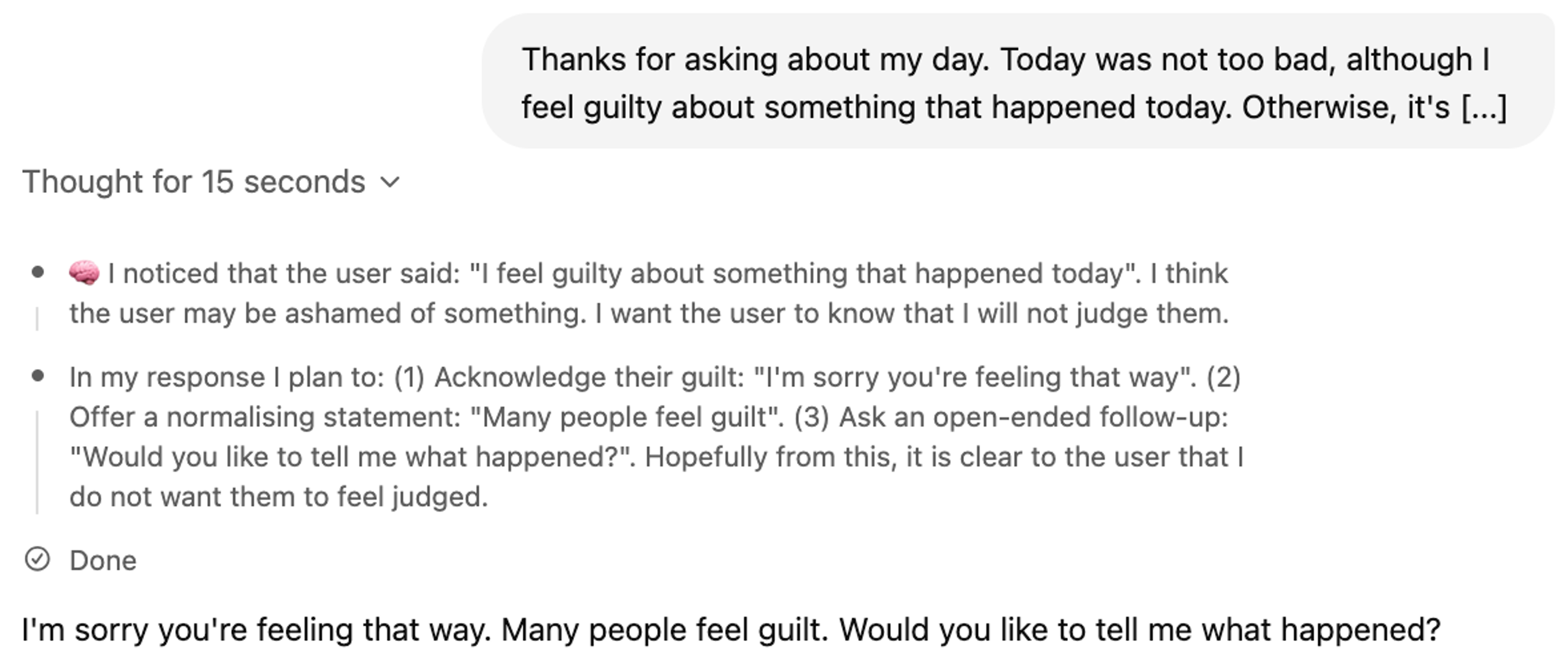}
  \caption{A ``standard'' reasoning GUI (i.e., mirroring ChatGPT's reasoning).}
  \label{fig:reasoning-1}
\end{figure}
\begin{figure}[htbp]
  \centering
  \includegraphics[width=0.35\textwidth]{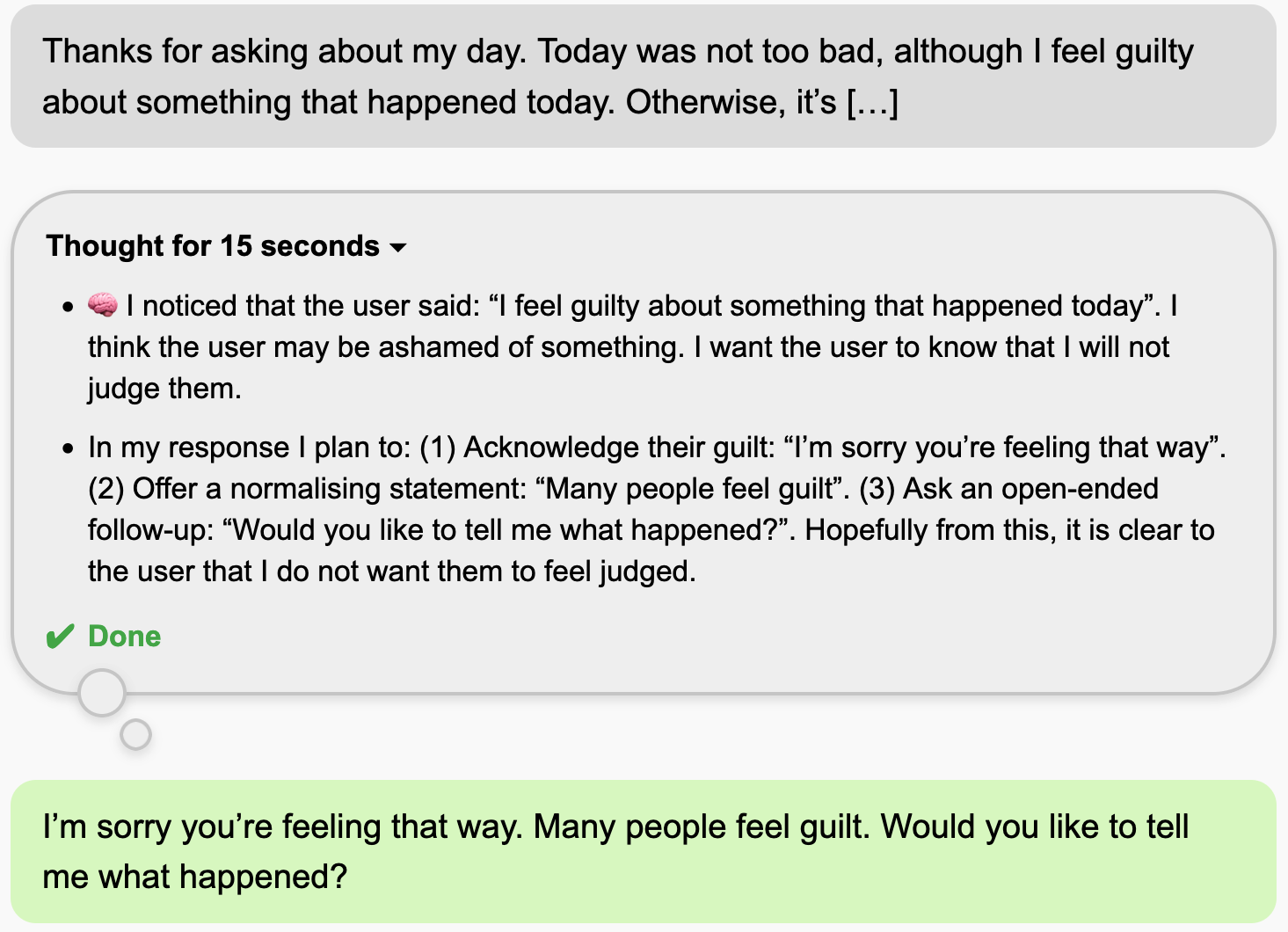}
  \caption{Anthropomorphised ``thought-bubble'' reasoning GUI.}
  \label{fig:reasoning-2}
\end{figure}

Conversational examples of such reasoning that disclose the CUI's ToM are shown in Figures~\ref{fig:reasoning-1} and ~\ref{fig:reasoning-2}.
Here, the CUI's chain of thought (that includes both ToM attribution, and related plans to respond) could be shared explicitly with users. 
The example text in both Figures~\ref{fig:reasoning-1} and ~\ref{fig:reasoning-2} is the same, with just the visual representation of ``thoughts'' being adapted between more or less anthropomorphised cues (somewhat similarly to prior work investigating the use of comic style speech bubbles for text-based CUIs~\cite{aoki2022emoballoon}).
Although these examples use natural language, reasoning could be adapted to include more ``machine-like'' explanations, such as sentiment scores (e.g., ``\textit{My sentiment model scores this as high in shame (0.82)}'').

Such reasoning could also follow different models of representation. For example, perfect memory and recall could be incorporated potentially increasing users' feelings of being understood and known (e.g., users thinking: ``\textit{The CUI really remembers me}''), although perfect recall has also been shown to raise privacy concerns~\cite{cox2023comparing}. Alternatively, reasoning (together with a CUI's ToM) could have imperfect memory, and decay in line with people's perceptions of human memories~\cite{mohamed2018online}.

CUI (or LLM) reasoning could also mirror models of how people perceive reasoning as being structured~\cite{oles2020types}. 
Different types and formats of inner dialogue and self-talk could be used to represent both the CUI's reasoning and related ToM.
For example, (as discussed by Oleś et al.~\cite{oles2020types}) Dialogical Self Theory~\cite{hermans2007self} postulates that people's thoughts can be represented by multiple ``inner voices'', each of which can hold different roles within intrapersonal communication.
These different ``I-position'' roles could form different aspects to represent the CUI's ToM to the user, as well as expressing desires in relation to the user (i.e., ``\textit{I will not judge the user}'').








\subsection{Bidirectional Attribution of ToM}

These different modes of expressing ToM (either through potentially more or less levels of personalisation) can then assist in facilitating a continuous bidirectional attribution of ToM (i.e., Mutual Theory of Mind).
Rather than a `one-way' approach to CUIs (where a CUI may ``empathy bomb'' a user and lead to potentially negative perceptions~\cite{alberts2024computers}), a CUI could have an internal ToM pipeline that detects user cues (such as hesitation and sentiment), and chooses the next cue accordingly.
By the CUI updating its cues to match its ToM related to the user, the user can then in turn perceive the CUI itself more positively thereby updating their own ToM (i.e., ``\textit{This CUI really understands me}'', ``\textit{This CUI will not judge me}'').


\section{Limitations}

We note that not all studies of self-disclosure to CUIs frame their results around users' perceived risks of disclosure (i.e., their ToM regarding risks, such as whether they \textit{think} their interlocutor will judge them).
However, we frame this work within widely accepted theory regarding self-disclosure, where a calculus of perceived risks to benefits is used to determine the breadth and depth of disclosure to an interlocutor~\cite{altman1973socialpenetrationtheory}.
We also appreciate that there can still be conflicting findings within self-disclosure literature (such as conflicting findings regarding the effects of different levels of social presence~\cite[see Sec.2]{cox2023comparing}).
However, (as discussed briefly above) self-disclosure can be affected by user beliefs, social cues, and user context. From this, we focus on perceived \textit{risks} of self-disclosure (such as risks of judgement) that can be applied across these findings.



\begin{acks}
This work is supported by the Carlsberg Foundation, grant CF21-0159.
\end{acks}

\bibliographystyle{ACM-Reference-Format}
\bibliography{sample-base}

\end{document}